\documentclass[prb,twocolumn,amsmath,amssymb]{revtex4}
\usepackage{graphicx}
\usepackage{dcolumn}
\usepackage{bm}
\usepackage{wasysym}
\usepackage{bbm}

\newcommand{\ie}{{\it i.e.},\ }

\newcommand{\grOh}{$\bm{\mathcal{O}}_{\!h}$\ }
\newcommand{\grO}{$\bm{\mathcal{O}}$\ }
\newcommand{\gridi}{$\{Id,i\}$\ }
\newcommand{\grGN}{$G_N$\ }
\newcommand{\grTN}{$T_N$\ }
\newcommand{\grPN}{$P_N$\ }
\newcommand{\Zdeux}{$\mathbb{Z}_2$\ }

\newcommand{\SUdeux}{$SU(2)$\ }
\newcommand{\SOtrois}{$SO(3)$\ }
\newcommand{\ganu}{$\Gamma_\nu$\ }
\newcommand{\ds}{$D_S$\ }
\newcommand{\csixv}{$C_{6v}$\ }
\newcommand{\ket}[1]{$|#1\rangle\ $}
\def\lsim{\mathrel{\rlap{\lower4pt\hbox{\hskip1pt$\sim$}}
    \raise1pt\hbox{$<$}}}
\def\gsim{\mathrel{\rlap{\lower4pt\hbox{\hskip1pt$\sim$}}
    \raise1pt\hbox{$>$}}}
\begin{document}

\preprint{cuboc}
 
\title{Twelve sublattice ordered phase in the \mbox{$\bm{J_1-J_2}$} model on the kagom\'e  lattice}
\author{J.-C.~Domenge}
\email{jcdomeng@lptl.jussieu.fr}
\author{P.~Sindzingre}
\author{C.~Lhuillier}
\affiliation{%
Laboratoire de  Physique Th\'eorique de la  Mati\`ere Condens\'ee,\ CNRS UMR
7600,\ Universit\'e  Pierre et Marie Curie,\  Bo\^ite Postale  121,\ 4
place Jussieu,\ 75252 Paris Cedex,\ France }%
\author{L.~Pierre}
\affiliation{%
Batiment G,\ Universit\'e Paris-X,\ Nanterre,\ 92001 Nanterre Cedex,\ France }%
\date{\today}
\begin{abstract}
Motivated by  recent experiments on an  \mbox{$S=1/2$} antiferromagnet
on the kagom\'e lattice,\ we investigate the Heisenberg \mbox{$J_1-J_2$}
model  with  ferromagnetic    $J_1$  and   antiferromagnetic  $J_2$.\\
Classically the  ground state displays N\'eel long-  range  order
with 12 noncoplanar  sublattices.\ The order parameter  has the
symmetry of a cuboctahedron,\    it fully breaks   \SOtrois as well
as  the spin flip  symmetry,\ and we expect  from the latter a
\Zdeux symmetry breaking pattern.\ As might be expected from the
Mermin-Wagner theorem in   two dimensions,\ the  \SOtrois symmetry
is  restored by thermal fluctuations  while  the \Zdeux symmetry
breaking persists   up to a finite temperature.\\  A complete
study of \mbox{$S=1/2$} exact spectra reveals that the classical
order subsists for  quantum spins in a finite range of
parameters.\ First- order spin wave calculations give the range of
existence of this phase and the renormalizations at \mbox{$T=0$}
of the order parameters associated to both symmetry breakings.\ 
This phase is destroyed by quantum fluctuations for a small but
finite $J_2/|J_1|\simeq3$,\  consistently with exact spectra
studies,\ which indicate a gapped phase.
\end{abstract}
\maketitle
\section{Theoretical and experimental issues}
Whatever the  nature  of the  spin,\ classical or  quantum,\ the
first neighbor Heisenberg  antiferromagnet  on the kagom\'e
lattice  fails to display N\'eel-like long-range order.\
Classically,\ it is characterized by  an  extensive
entropy~\cite{e89,chs92} at \mbox{$T=0$}.\  Quantum mechanically
the spin-1/2 system has an exceptionally large density of low
lying   excitations~\cite{lblps97,web98} reminiscent      of  the
classical  extensive   entropy.\  It  is still   debated   whether
and eventually how    this   degeneracy   is   lifted  in    the
quantum limit~\cite{ml05,ba04}.\\ An essential issue concerns the
influence of perturbations:\  classically the  effect of  a second
neighbor coupling $J_2$ has been very early  studied by Harris and
co-workers~\cite{hkb92}.\  They showed that  an infinitesimal
$J_2$ is  sufficient to drive the system toward an ordered phase
with the   three  spins around  a triangle pointing 
$120^\circ$    from    each  other.\ Antiferromagnetic
second-neighbor coupling (\mbox{$J_2>0$}) favors the \mbox{${\bf
q}={\bf  0}$} N\'eel order  of this  pattern  on the Bravais
lattice,\   whereas  there   are   nine   spins per unit  cell for
\mbox{$J_2<0$}     (\mbox{${\bf q}=\sqrt{3}\times\sqrt{3}$}
order).\   The effect    of Dzyaloshinsky-Moriya interactions has
also been analyzed~\cite{ecl02}.\  To our  knowledge  the reduction
of the order  parameter by quantum fluctuations  has only been
studied through exact diagonalizations \cite{waldtmannphDthesis}.\ 
This approach points to an immediate transition from the "disordered phase"
 at the pure $J_1>0$ point,\  to the semiclassical N\'eel phases.   \\

Up until now  the \mbox{$J_1-J_2$} model on   the kagom\'e lattice
has only  been studied for antiferromagnetic $J_1$.\ Many magnetic
compounds~\cite{r94,rhw00, mklmch00,bmcbbabh04}  with this
geometry   have been studied so far,\  but most of them have spin
\mbox{$S=3/2$}.\   A few compounds with \mbox{$S=1/2$} Cu ions have
recently been synthetized~\cite{h01,knisu02,mbp01}.\ None of them
can be described by a pure isotropic first neighbor
antiferromagnetic Heisenberg model.\ Recent experimental work on
an organic compound with copper   ions on  a kagom\'e
lattice~\cite{nkh04}   gives indication of competing ferromagnetic
and  antiferromagnetic  interactions.\\
It is thus the purpose of the present work to extend the previous
study of the \mbox{$J_1-J_2$} model to ferromagnetic nearest
neighbor coupling (\mbox{$J_1<0$}).\  The Hamiltonian reads as
\begin{equation}
{\cal  H} = J_1\!\!\sum_{<i,j>}  {\bf S}_i\cdot{\bf S}_j + J_2\!\!\sum_{<<i,k>>}
{\bf S}_i\cdot{\bf S}_k, \label{H}
\end{equation}
where the first and second sums run,\ respectively,\ over pairs of nearest
neighbors   \mbox{$<i,j>$}   and          next-nearest       neighbors
\mbox{$<<i,k>>$}.\\ For a pure ferromagnetic
$J_1$  coupling the system is indeed  in a ferromagnetic phase.\ For a
pure antiferromagnetic $J_2$ interaction the  model reduces
to   three   decoupled    kagom\'e   lattices  with    antiferromagnetic
interactions  and  has thus an   extensive entropy in  the classical
limit.\   The behavior of   the model between  these  two limits is the
object of the  present   study.  \\ In   Sec.~\ref{sec:classic} the
classical ground state  of  the Hamiltonian~(\ref{H}) is  investigated
and the  phase diagram of the model  is given in the classical limit.\
For competing interactions
\mbox{$J_1<0$} and \mbox{$J_2>0$} the model  exhibits an ordered phase
with 12 sublattices that fully breaks \SOtrois as  well as a discrete
symmetry  (chiral  symmetry breaking).\ We  show that  contrary to the
N\'eel  order,\ which  breaks a   continuous  symmetry and therefore  is
destroyed down   to  infinitesimal  temperatures,\  the  chiral  order
survives thermal  fluctuations  and undergoes  a phase  transition  at
finite  temperature.\\    In Sec.~\ref{sec:quantum}  we  study  the
\mbox{$S=1/2$} quantum model using exact diagonalizations and show the
premise of the semi-classical ordering on samples up to 36 spins.\\ In
Sec.~\ref{sec:sw} we study   the effect of long  wavelength quantum
fluctuations  on this    semi-classical    order in    the   spin-wave
approximation.\ It  appears  that  the  twelve sublattice  N\'eel  order
survives quantum  fluctuations in an  extended range of parameters but
is destroyed for a non zero \mbox{$J_2>0$} ($ J_2/|J_1|\gsim 3$).\\ In
Sec.~\ref{sec:gappedphase} we  show that exact  diagonalizations in
this range of parameters indeed point to (a) gapped phase(s).
\section{Classical approach}
\label{sec:classic}
In this section  we restrict ourselves  to classical spins:\ the spins
${\bf S}_i$    are usual unit vectors   living  in a three-dimensional
space.
\subsection{Ground state for \mbox{$\bm{J_1<0}$} and \mbox{$\bm{J_2>|J_1|/3}$}}
\begin{figure}[h]
\resizebox{.2\textwidth}{!}{\includegraphics*[0cm,0cm][8.5cm,7.5cm]{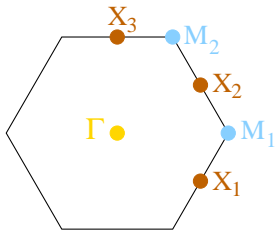}}
\caption{(Color online)\label{fig:1zB} The first Brillouin zone of the
kagom\'e lattice with its points of high symmetry:\ the center of
zone $\bm{\Gamma}$ \mbox{(${\bf q}={\bf 0}$)},\ the two zone corners
$\bm{M}_1$ and $\bm{M}_2$,\ and the three edge centers $\bm{X}_1$,\ $\bm{X}_2$,\ and
$\bm{X}_3$.}
\end{figure}

\begin{figure}[h]
\resizebox{.45\textwidth}{!}{\includegraphics*[0cm,0cm][18cm,7.5cm]{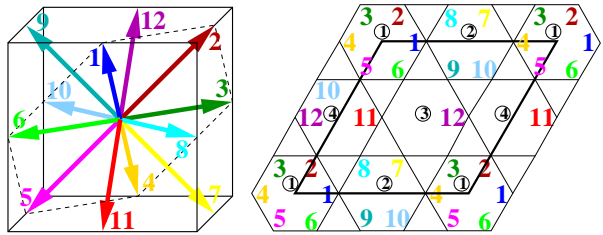}}
\caption{(Color online)\label{fig:cuboc} Classical order parameter
for \mbox{$J_1<0$} and     \mbox{$J_2>|J_1|/3$}  with the
associated  lattice  symmetry breaking.\  The 12 sublattices
and  the 4 different types  of hexagons are  numbered.\  Also
shown is  the  plane containing the six spins of hexagon 1
(dashed).}
\end{figure}

\begin{figure}[h]
\resizebox{.3\textwidth}{!}{\includegraphics*[0cm,0cm][8.5cm,8.5cm]{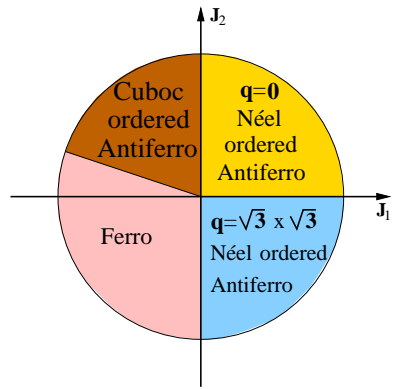}}
\caption{(Color online)\label{fig:diagphase} Classical phase
diagram of the Heisenberg \mbox{$J_1-J_2$} model on the kagom\'e
lattice at \mbox{$T=0$}.\ The transition between the ferromagnetic
state and the antiferromagnetic {\it cuboc} state occurs at
\mbox{$J_2=-J_1/3$}.}
\end{figure}

To    investigate  the  nature   of  the     ground    state  of   the
Hamiltonian~(\ref{H}),\ we   first  Fourier transform  it  to   find its
lowest-energy modes.\   The  kagom\'e  lattice having  three   sites per
Bravais cell,\  we    get  three  branches.\  For   \mbox{$J_1<0$}   and
\mbox{$J_2<0$} we find a single minimum at
\mbox{${\bf q}={\bf  0}$} corresponding  to the expected ferromagnetic
ground state.\ Upon increasing \mbox{$J_2>0$},\ we find three degenerate
minima  at  the edge centers   of the first   Brillouin zone,\ the
classical transition occuring at \mbox{$J_2=|J_1|/3$}.\ The three modes
\mbox{$\bm{q}=\bm{X}_{1,2,3}$} (Fig.~\ref{fig:1zB}) are the only solution
as long as \mbox{$J_2>|J_1|/3$} and in the limit of pure \mbox{$J_2>0$}
one recovers the  flat zero-energy branch  of the  pure \mbox{$J_1>0$}
case:\ we  then   have three  decoupled kagom\'e lattices   with nearest
neighbor coupling $J_2$.
In  the  \mbox{$J_1<0$},\  \mbox{$J_2>|J_1|/3$}  region,\ the   unit cell
compatible  with the three  edge centers  contains 12 sites and the
direct minimization of~(\ref{H}) for small samples of size multiple of
12  indeed  reveals  a N\'eel   long-range order   with  12 noncoplanar
sublattices  pointing  toward   the 12  centers  of   edges of  a cube
(Fig.~\ref{fig:cuboc}).
The apparent complexity of this structure is somewhat lightened when
one considers the six spins around  an hexagon:\ they  lie in the same
plane and make an angle $\pi/3$  with their nearest neighbors.\ We thus
have  four hexagons in  the unit  cell  defining four different planes
oriented like  the   faces  of a  tetrahedron.\ The    polyedron whose
vertices coincide  with the directions  of the sublattices  is named a
cuboctahedron,\ and we will refer to that order parameter as the
\mbox{\it cuboc} phase in the following.\ Consistently,\ Monte Carlo
simulations with   a Metropolis algorithm   reveal a  local  \mbox{\it
cuboc} N\'eel order  with fluctuations increasing with temperature.\\ We
now have a  complete  picture of the   classical phase diagram  of the
Hamiltonian~(\ref{H})  at \mbox{$T=0$}  in  the  entire \mbox{$J_1-J_2$}
plane (Fig.~\ref{fig:diagphase}).
\subsection{Discrete symmetry breaking}
\begin{figure}[h]
\resizebox{.4\textwidth}{!}{\includegraphics*[0cm,0cm][17cm,9.5cm]{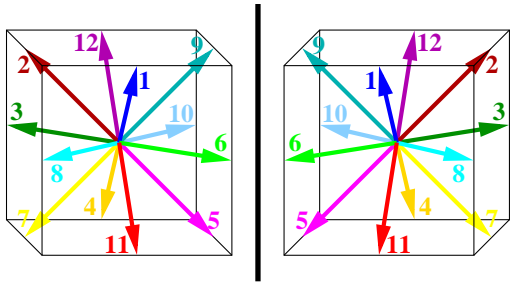}}
\caption{(Color online)\label{fig:cubIcub} An order parameter and
its  image by a  spin flip.\ Clearly,\ after  a global rotation of the
sublattice directions,\ the two  order  parameters are related one  to
the other by a mirror symmetry.}
\end{figure}

\begin{figure}[h]
\resizebox{.4\textwidth}{!}{\includegraphics*[0cm,0cm][15cm,10cm]{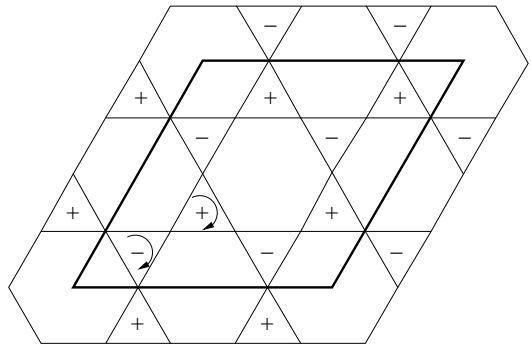}}
\caption{\label{fig:chitri} The normalized scalar chirality
$\sigma_\bigtriangleup$ computed on the order parameter of
Fig.~\ref{fig:cuboc}.\ The spin flip trivially permutes the $+$
and $-$.}
\end{figure}

\begin{figure}[h]
\resizebox{.4\textwidth}{!}{\includegraphics*[1.3cm,1.2cm][20cm,20cm]{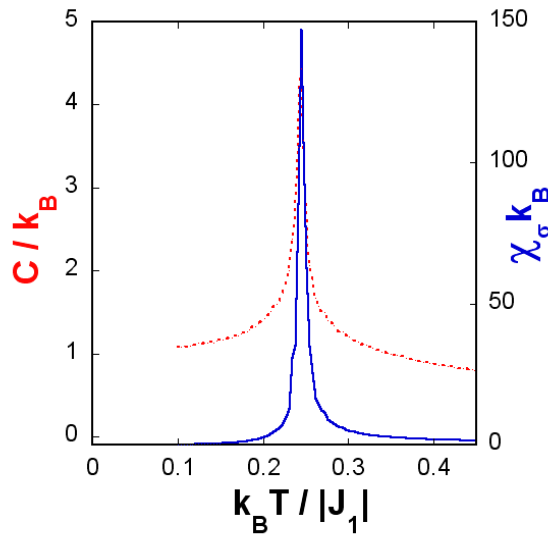}}
\caption{(Color online) \label{fig:cvkhi} Specific heat  \mbox{$C/k_B$} 
(red dashed  line)   and    chiral  susceptibility  \mbox{$k_B\chi_\sigma$}
(blue solid line) versus temperature  on a  1200  spins sample for
$J_2/|J_1|=0.5$:\ both quantities diverge at
\mbox{$T_c=0.24$},\ simultaneously with the vanishing of the chiral order
parameter $m_\sigma$ (not shown).}
\end{figure}

It is clear that $O(3)$ is fully broken  in the \mbox{\it cuboc} phase
at \mbox{$T=0$}:\ the point    group symmetry of a cuboctahedron    is
simply that of the cube,\ \ie \mbox{\grOh=\ \grO$\!\!\times$\gridi},\ where
\grO  is the octahedral group   containing the 24  rotations leaving a
cube  or an   octahedron invariant,\  and  \mbox{$Id$}   and $i$ are,\
respectively,\ the  identity and the  spin inversion,\ or spin flip.\\
An important  result arises when one  considers the action of the spin
flip alone.\  It is clear that $i$  acting on a cuboctahedron takes it
onto another cuboctahedron but the    labels of the two   cuboctahedra
cannot be made to  coincide by means  of a global rotation.\  Namely,\
the order parameter   we obtain is  the  mirror-symmetry image of  the
previous one (Fig.~\ref{fig:cubIcub}).
The order  parameter  in the \mbox{\it cuboc}  phase  thus breaks  the
spin flip   symmetry, and  we are able    to divide  the  ground states
manifold  into two  classes.\  This makes us   expect a transition  at
finite temperature associated  with this \Zdeux symmetry  breaking.\\ To
show  that  this is indeed the  case,\  we define a variable
labeling the two classes  of ground states.\ Consider  the normalized
scalar chirality on a triangle,\ namely
\mbox{$\sigma_\bigtriangleup=\frac{(\bm{S}_i\land\bm{S}_j)\cdot\bm{S}_k}{|(\bm{S}_i\land\bm{S}_j)\cdot\bm{S}_k|}$}
with   \mbox{($i$,$j$,$k$)} labeling    the three  sites  clockwise.\
An inspection of the order parameter of Fig.~\ref{fig:cuboc} reveals that
\mbox{$\sigma_\bigtriangleup$} is alternatively +1  on upward triangles
and -1 on  downward  triangles (Fig.~\ref{fig:chitri}).\ We  naturally
define the alternate scalar chirality as
\begin{equation}
m_\sigma=\frac{3}{2N}\sum_\bigtriangleup(-1)^{\alpha_\bigtriangleup}\sigma_\bigtriangleup,
\end{equation}
where the sum runs  over the $2N/3$  triangles of the kagom\'e lattice,\
and $\alpha_\bigtriangleup$ is,\ respectively,\  0 and 1 on  upward and downward triangles.\
The  spin flip  trivially changes the  sign  of
$\sigma_\bigtriangleup$ so  that \mbox{$m_\sigma=\pm1$} at zero
temperature,\ depending   on the class of  the order  parameter.\
Hence,\ $m_\sigma$ is  the order parameter associated with the
spin flip symmetry breaking.\\ Monte Carlo  simulations have been  performed
on  samples  of up to 1200 spins: they show that  $m_\sigma$ 
vanishes at finite temperature while the  associated chiral
susceptibility,\  defined as
\begin{equation}
k_B\chi_\sigma=\frac{2N}{3T}(<m_\sigma^2>-<|m_\sigma|>^2),
\end{equation}
and the specific heat,\  seemingly both diverge (Fig.~\ref{fig:cvkhi}).\
Typical simulations involved $10^6$ Monte Carlo steps per spin.\ The first
results indicate that  the transition is  not in the  two-dimensional
Ising  universality  class.\  The  complexity  of  the  global set  of
excitations that does not reduce to those of the above-mentionned $\sigma_\bigtriangleup$
variable are  probably at the   origin of  a more  complex  behavior,\
presumably   a  weak first-order  phase  transition~\cite{dmt04}.\ The
complete study  of this classical phase  transition will  be published
elsewhere~\cite{dlv05}.
\section{Spin-1/2 model}
\label{sec:quantum}
We now turn to the spin-1/2  quantum model.\ Considering the classical
analysis,\ the question  is whether  quantum fluctuations are   strong
enough to wipe  out  the classical order.\ If  so,\  we end up  with a
purely   quantum   phase   such  as  a    valence   bond solid   or  a
RVB liquid~\cite{ml05}.\ If not,\ this means that quantum fluctuations
merely  dress the classical   order   parameter,\ reducing the    mean
sublattice magnetization while preserving its symmetries.\ Such states
may thus be referred to as {\it semiclassical} states.
\subsection{\`A la N\'eel $\bm{SU(2)}$ symmetry breaking}
\label{subsec:alaneel}
The exact spectra of  finite size samples have  already been shown to be a
powerful   tool   to find  eventual    N\'eel  orders  in   quantum spin
systems~\cite{blp92,bllp94}.\ Of  course  they  allow  for  the  exact
computation  of   the square  of the  sublattice  magnetization or any
relevant structure factor.\ But   mostly,\ while the extensive  use of
the  symmetries     of the Hamiltonian  is   compulsory   in  order to
diagonalize large enough samples (typically  36 spins today),\ it also
gives  a clear  signature of \mbox{$SU(2)$-breaking}  phases,\ even on
small samples spectra.\\  Namely,\ for an \SUdeux symmetry  breaking
we expect a large  set of low-lying  eigenstates of  the Hamiltonian,\
with different total    spin   $S$ values,\  to   collapse   onto  the
ground state  when the size of  the sample \mbox{$N\to\infty$}.\ These states have
been  called  the {\it quasidegenerate   joint states} (QDJS) and they
have       been    recently         observed   experimentally       on
nanomagnets~\cite{w05}.\  If   the semiclassical picture  is valid,\
these QDJS   are expected  to have  energies  well below   the  magnon
excitations  and  to scale  as  \mbox{$S(S+1)/N$},\  as  expected for  a
quantum top.\ Exact spectra are thus displayed versus
\mbox{$S(S+1)$}  and the QDJS are often   referred to as the Anderson's
tower of states~\cite{bllp94}.\ To  have a true \SUdeux symmetry breaking
in the thermodynamic limit,\ the QDJS should have total spins up to
\mbox{$S\sim\mathcal{O}(\sqrt{N})$}.\ The breaking of the \SUdeux
symmetry then  occurs with all the QDJS   collapsing onto the absolute
ground  state  like  $1/N$ when  \mbox{$N\to\infty$},\ \ie    faster than the
softest magnons,\ whose energies scale as $1/\sqrt{N}$,\ defining the
\mbox{\it \`a la N\'eel} \mbox{$SU(2)$-symmetry} breaking scheme~\cite{bllp94}.\
This result  explains why it is  numerically more favorable to look at
the QDJS,\  since order parameters  only scale as $1/\sqrt{N}$.\\ In the
thermodynamic  limit,\ the  ground   state is a  superposition  of  an
infinite number of  QDJS  with different  $S$  values,\ which  clearly
breaks  \SUdeux.\\ The crucial point is  that  in each spin sector the
number and symmetries  of  the QDJS  are exacly  determined from group
representation theory~\cite{bllp94,lblps97}  by  the symmetry   of the
expected ground state.
\subsection{Determination of the QDJS}
If the thermodynamic ground state exhibits a semiclassical
\mbox{12-sublattice} N\'eel order then the QDJS,\ if ever they exist,\
should   be   of symmetry    compatible    with  both  those   of  the
Hamiltonian~(\ref{H}),\ since they are eigenstates,\ and those of
the \mbox{\it cuboc} phase,\ \ie  \grOh.\\ A classical result of group
theory is that the  number of such states  is completely determined by
the structures  of the two groups.\  Indeed,\ if we restrict ourselves
to the rotational symmetry breaking,\ we see that the original
\SOtrois symmetry  of~(\ref{H}) is reduced to its  subgroup \grO in the
N\'eel-ordered   ground  state.\\  Thus,\ while   \ds  is an irreducible
representation (IR)  of \SOtrois of spin  $S$,\ it is an  \mbox{\it a
priori} reducible representation of  \grO that one can  decompose onto
the five IRs \ganu of \grO according to
\begin{equation}
D_S=\sum_{\nu=1}^5n_\nu(S)\,\Gamma_\nu,
\label{dec}
\end{equation}
with
\begin{equation}
n_\nu(S)=\frac{1}{24}\sum_{g\in\mathcal{O}}\chi_\nu^*(g)\chi_{_S}(g),
\label{nnus}
\end{equation}
where \mbox{$\chi_\nu(g)$} and \mbox{$\chi_{_S}(g)$} are the characters of
\mbox{$g\in\bm{\mathcal{O}}$} in the IR \ganu of \grO,\ and in the IR \ds of
\SOtrois,\ respectively.\ The character table of \grO is given in
Table~\ref{tab:IR}                                                 and
\mbox{$\chi_{_S}(g)=\frac{\sin((2S+1)\theta/2)}{\sin(\theta/2)}$},\   with  $\theta$ the
angle  of the rotation  $g$.\  For completeness  we give the  explicit
decomposition~(\ref{dec})    for    spins up    to    \mbox{$S$=6}  in
Table~\ref{tab:decex}.
\begin{table}[h]
\caption{\label{tab:IR} Character table of \grO.\ Its irreducible
representations are  named both with  the usual nomenclature  and with
the numbering \mbox{$1\leq\nu\leq5$} used throughout  the article.\ Also shown
is   the character table of  the  group \gridi.\ The character
table  of the whole  \grOh  group is  obtained  by forming the  direct
product of the two tables.\\}
\begin{tabular}{lcr}
\parbox{.3\textwidth}{
\framebox{\begin{tabular}{r|rrrrr||c}
$\bm{\mathcal{O}}$     & $Id$  & 8$C_3$  & 3$C_2$ & 6$C_2$ & 6$C_4$ & $\nu$\\\hline
$A_1$ &  1    &    1   &   1   &   1   &   1   & 1\\
$A_2$ &  1    &    1   &   1   &  -1   &  -1   & 2\\
$E$  &  2    &   -1   &   2   &   0   &   0   & 3\\
$T_1$ &  3    &    0   &  -1   &  -1   &   1   & 4\\
$T_2$ &  3    &    0   &  -1   &   1   &  -1   & 5\\
\end{tabular}}}
&\parbox{.1\textwidth}&
\parbox{.1\textwidth}{
\framebox{\begin{tabular}{c|rr}
$\{Id,i\}$     & $Id$ & $i$\\\hline
$\Gamma_e$ & 1  & 1\\
$\Gamma_o$ & 1  & -1\\
\end{tabular}}
}
\end{tabular}
\end{table}\\
The decomposition~(\ref{dec}) directly gives the number of states that
belong both to \ds and \ganu,\ \ie that are compatible with both the
\SOtrois and \grO symmetries,\ as required for the QDJS.\ It should be
emphasized that  for a given  $S$ value the number  of such  states is
\mbox{$n(S)=\displaystyle{\sum_{\nu=1}^5n_\nu(S)\dim\Gamma_\nu}=2S+1$},\ as expected
for a complete \SOtrois  breaking.~\cite{bllp94}\\ We stress that  (\ref{dec}) relies
only  on  group  theory   and that  it   makes  no  reference  to  the
representation space\footnote{In our  problem the representation space
is
\mbox{$\mathcal{E}=\displaystyle{\bigotimes_{i=1}^{12}}\mathcal{D}_{N/24}$}.\
For large values of $S$ the subspace  of $\mathcal{E}$ with total spin
$S$ has a dimension smaller than \mbox{$2S+1$} and  can only display a
limited  set of the  states appearing in the decomposition~(\ref{dec}).\
However,\ at least  up to \mbox{$S<\sqrt{N}$},\  it can be verified by
direct  inspection  that the  number  of states  in $\mathcal{E}$ with
total spin $S$ is much larger than \mbox{$(2S+1)$}.}.\\
\begin{table}[h]
\caption{\label{tab:decex}The decomposition~(\ref{dec}) explicited for
\mbox{$S\leq6$}.\\}
\framebox{
\begin{tabular}{ccccccccccc}
$D_0$&=&  $A_1$ & & & &  & & &  &\\ $D_1$&=& $T_1$ & &  & &  & & & &\\
$D_2$&=& $E$ & + & $T_2$ & & & & & &\\ $D_3$&=& $A_2$ &  + & $T_1$ & +
& $T_2$ & & & &\\ $D_4$&=& $A_1$ & + & $E$ &  + & $T_1$  & + & $T_2$ &
&\\ $D_5$&=& $E$ & + & $2T_1$ & + & $T_2$ & & & &\\ $D_6$&=& $A_1$ & +
& $A_2$ & + & $E$ & + & $T_1$ & + & $2T_2$\\
\end{tabular}}
\end{table}
Now,\ if we are to find the  total content of  each spin sector of the
Anderson's tower of states,\ we should treat the whole
\grOh group,\ not limiting   ourselves  to the rotational 
symmetry-breaking \mbox{$SO(3)\to\bm{\mathcal{O}}$} as in~(\ref{dec}).\\
This is particularly simple since \grOh is the direct product of
\grO with the group \gridi,\ and the spin flip being also a symmetry
of  the Hamiltonian,\ there  is  no compatibility issue here.\  Thus,\
everytime an  IR \ganu   appears  in~(\ref{dec}) we actually   get two
copies  of   it  associated to  the   two  IRs  of  the \gridi   group
(Table~\ref{tab:IR}).\ Since these two IRs differ only in their parity
under the spin flip   operation,\  which itself transforms any   order
parameter into  its \mbox{\Zdeux$\!\!$ image},\  this double \mbox{\it
quasidegeneracy}    is  clearly  reminiscent  of  the  \Zdeux symmetry
breaking observed classically.\\ We thus  have formally determined the
number and symmetries  of the QDJS appearing in  each spin sector of
the  tower  of  states.\  However,\ for   a  given  total spin,\ their
symmetries   are given in   terms  of the  IRs   of \grOh while  exact
diagonalizations provide eigenstates   of  given symmetry   under  the
lattice symmetry group.\\ It thus remains to map the IRs of \grOh onto
those of the lattice  symmetry group,\ namely \mbox{$G_N=T_N\land  P_N$},\
where  \grTN contains  the $N/3$ translations   by  a Bravais  lattice
vector and \grPN is the point group of the sample (in general,\
\grPN is  a subgroup of  \csixv,\  the point group  of the
infinite kagom\'e  lattice).\\ Such a mapping  clearly exists  since the
labeling of the 12 vertices of the  cuboctahedron induces a labeling
of the lattice (Fig.~\ref{fig:cuboc}).\ Applying an element of
\grOh to a cuboctahedron means permuting its  12 labels,\ which itself
is equivalent  to a lattice transformation.\\ We  thus  have a mapping
between \grOh and some elements of \grGN.\ Note that while the mapping
between the group elements is  not necessarily \mbox{one-to-one},\ and
in fact it is a \mbox{one-to-many} mapping for all the elements of the
subgroup  \grO,\  the resulting  mapping between the  IRs  of  the two
groups  {\it    is}  \mbox{one-to-one},\  as   is   explicited  in
Table~\ref{tab:map}.\\ A notable exception is  the spin flip,\ which  is
exactly mapped onto  the rotation of the  lattice  $R_\pi$ by  angle $\pi$
around  the center  of  an empty  hexagon.\ Hence,\   the parity  of  an
eigenstate of the  Hamiltonian~(\ref{H}) under $R_\pi$ is directly equal
to its parity under the spin flip.
\begin{table}[h]
\caption{\label{tab:map} \mbox{One-to-one} mapping between the IRs of
\grOh and those of \grGN.\ We considered a sample having the three
edge centers in its first Brillouin zone and the full \csixv point symmetry.\
\csixv is generated by  $R_{2\pi/3}$,\ the rotation of the lattice
by angle $2\pi/3$ around an empty hexagon,\ $R_\pi$ and $\sigma$,\ a reflection
whose axis is a  diameter  of an empty  hexagon,\  so that the IRs  of
\csixv may  be  labeled   by the associated three  quantum numbers
$\mathcal{R}_{2\pi/3}$,\ $\mathcal{R}_\pi$,\ and $\sigma$.\ The  IRs of \grTN are
labeled by the ${\bm k}$ vector of the first Brillouin zone.\\}
\begin{tabular}{cc}
\parbox{.2\textwidth}{
\framebox{\begin{tabular}{r|ccr||c}
$\bm{\mathcal{O}}$     & $\bm{k}$ & $\mathcal{R}_{2\pi/3}$ & $\sigma$ & $\nu$\\ \hline
$A_1$ & $\bm{0}$ & 1 & 1 & 1\\
$A_2$ & $\bm{0}$ & 1 & -1 & 2\\
$E$  & $\bm{0}$ & $j,j^2$ & & 3\\
$T_1$ & $\bm{X}_{1,2,3}$ & & -1 & 4\\
$T_2$ & $\bm{X}_{1,2,3}$ & & 1 & 5\\
\end{tabular}}}
&
\parbox{.2\textwidth}{
\framebox{\begin{tabular}{c|r}
$\{Id,i\}$     & $\mathcal{R}_{\pi}$\\ \hline
$\Gamma_e$ &  1              \\
$\Gamma_o$ & -1              \\
\end{tabular}}
}
\end{tabular}
\end{table}
Thus,\ if ever the thermodynamic ground state has the classical
\mbox{12-sublattice} structure,\   we are  able   to   find  the number   and
degeneracies of the expected QDJS using Table~\ref{tab:decex} and the
\mbox{one-to-one} mapping between the IRs of \grOh and those of
\grGN (Table~\ref{tab:map}).\\ However,\ there is still one subtlety.\
Indeed,\ in order not to artificially frustrate the
\mbox{12-sublattice} order,\ we choose samples containing multiples of 12
spins,\  \ie  \mbox{$N$=12},\ 24,\ and   36  spins.\ The  representation
spaces  of these three  samples  have  different properties since  the
total  spin  on each  sublattice is   $N/24$,\  which may be  integer or
half-integer.\\ To be more specific,\ we  want to write down the matrix
\mbox{$\hat{U}(g)\in    SU(2)$} associated   with  a  particular  rotation
\mbox{$g\!\!\in$\grO}  that acts on the  wave function  \ket{cuboc} of a
\mbox{{\it cuboc}-ordered}  state.\ The  Hilbert  space that  contains
such states is a subspace of
\mbox{$\displaystyle{\bigotimes_{i=1}^{12}}\mathcal{D}_{N/24}$},\  where  $\mathcal{D}_{N/24}$
is the Hilbert space of one spin $N/24$.\ Thus,\ a natural choice for
\mbox{$\hat{U}(g)$} would be the tensor product of 12
\mbox{$\hat{U}_{N/24}(g)$ matrices},\ each one of them representing $g$
in    $\mathcal{D}_{N/24}$.\\ Now,\ we   know    that in  each  subspace
$D_{N/24}$,\ if $N/24$ is a half-integer,\
\mbox{$\hat{U}_{N/24}(g)$} and \mbox{$-\hat{U}_{N/24}(g)$} are equally suitable
choices,\ and we cannot decide between them other than arbitrarily,\ due
to   the  double connectedness of \SOtrois.\\   Let us  choose  12 such
matrices anyway  and form  their tensor  product \mbox{$\hat{U}(g)$}.\
When acting on a  particular order parameter,\ named  \ket{cuboc},\ we
get
\begin{equation}
\hat{U}(g)|cuboc\rangle=\varphi(g)|cuboc'\rangle,
\end{equation}
where \ket{cuboc'} represents the order parameter obtained by applying
the  global  rotation    \mbox{$g\!\!\in$\grO}  on  \ket{cuboc},\    and
\mbox{$\varphi(g)$}  is an overall  phase factor  that we cannot  get rid of
since it  embeds the  arbitrariness of our  choice  of the  matrices
\mbox{$\hat{U}_{N/24}(g)$}.\\ Recall,\ however,\ that     \mbox{$\hat{U}(g)$}
acts in a subspace of
\mbox{$\displaystyle{\bigotimes_{i=1}^{12}}\mathcal{D}_{N/24}$},\  which   is known from
spin algebra  to  contain states  with  integer spins  only,\ whatever
$N/24$,\  integer,\ or  half-integer.\  Hence  \mbox{$\hat{U}(g)$}  is
always a true representation of $g$,\ \ie  no ambiguity should remain in
it and consequently  in \mbox{$\varphi(g)$}.\ Thus the group law should
be exactly   verified    by \mbox{$\varphi(g)$},\ which   is  then   simply  a
one-dimensional,\ thus irreducible,\ representation of
\grO.\\ Direct computation  for \mbox{$N$=12},\ 24,\ and 36 spins indeed
shows that \mbox{$\varphi(g)=\chi_{\nu_0}(g)$},\ where \mbox{$\nu_0$=1} for
\mbox{$N$=24} spins,\ and
\mbox{$\nu_0$=2} for \mbox{$N$=12} and 36 spins.\\ Finally,\ to  embed this
subtlety in the decomposition~(\ref{dec}),\ one simply has to permute
\mbox{$A_1\leftrightarrow A_2$} and \mbox{$T_1\leftrightarrow T_2$} in Table~\ref{tab:decex} for
\mbox{$N$=12} and 36 spins,\ as can be seen directly on the character
table  of    \grO  (Table~\ref{tab:IR}).\ Note    that  no additionnal
ambiguity arises from the consideration of the  \gridi part of \grOh,\
since the whole discussion relies on the double connectedness of
\SOtrois that is irrelevant here.
\subsection{Analysis of exact \mbox{$\bm{N}$=36} spectrum}
\begin{figure}[h]
\resizebox{.4\textwidth}{!}{\includegraphics*[0cm,0cm][18cm,12.2cm]{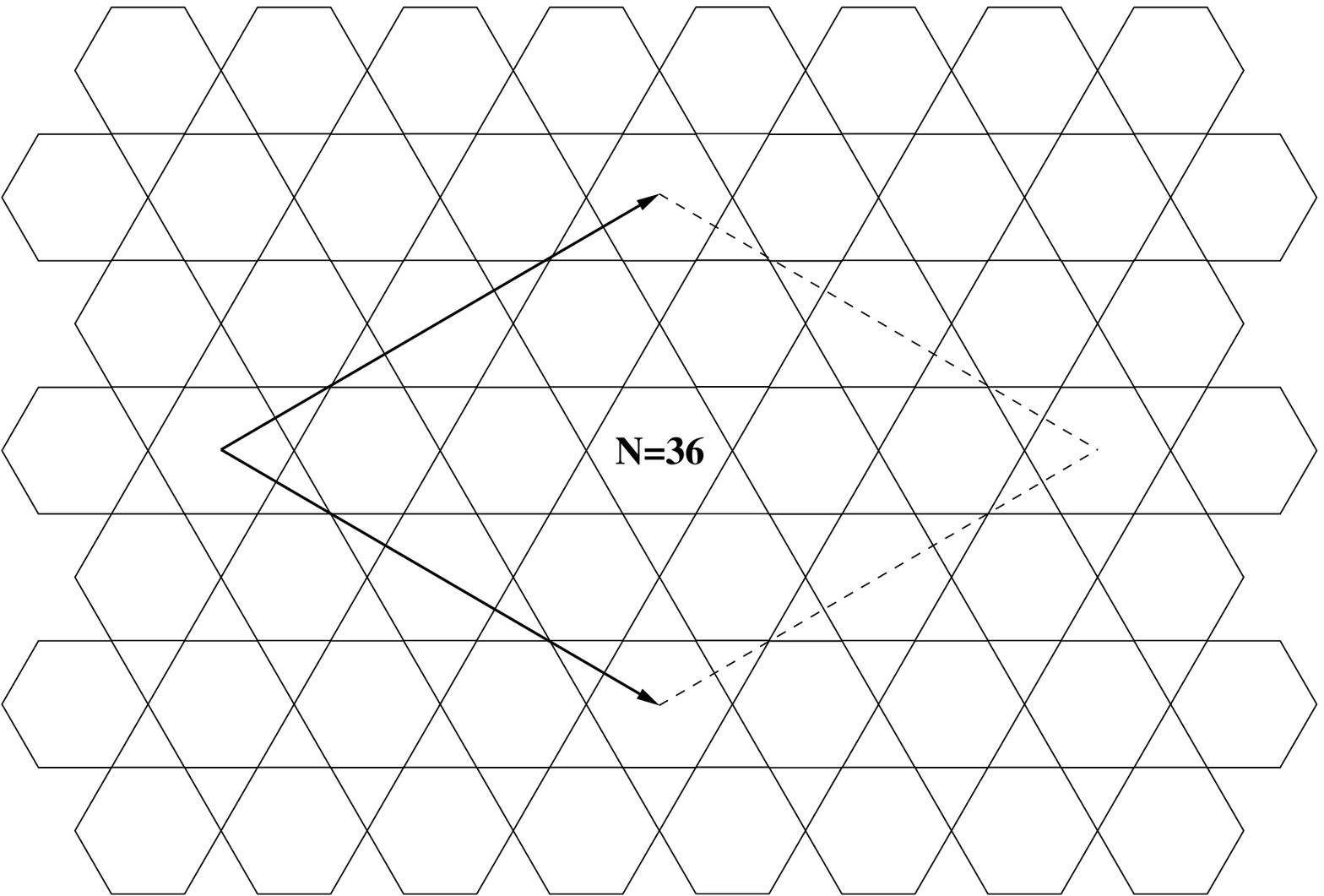}}
\caption{\label{fig:samples} \mbox{$N$=36} sample.}
\end{figure}

\begin{figure}[h]
\resizebox{.4\textwidth}{!}{\includegraphics*[0.5cm,7.5cm][20cm,24.2cm]{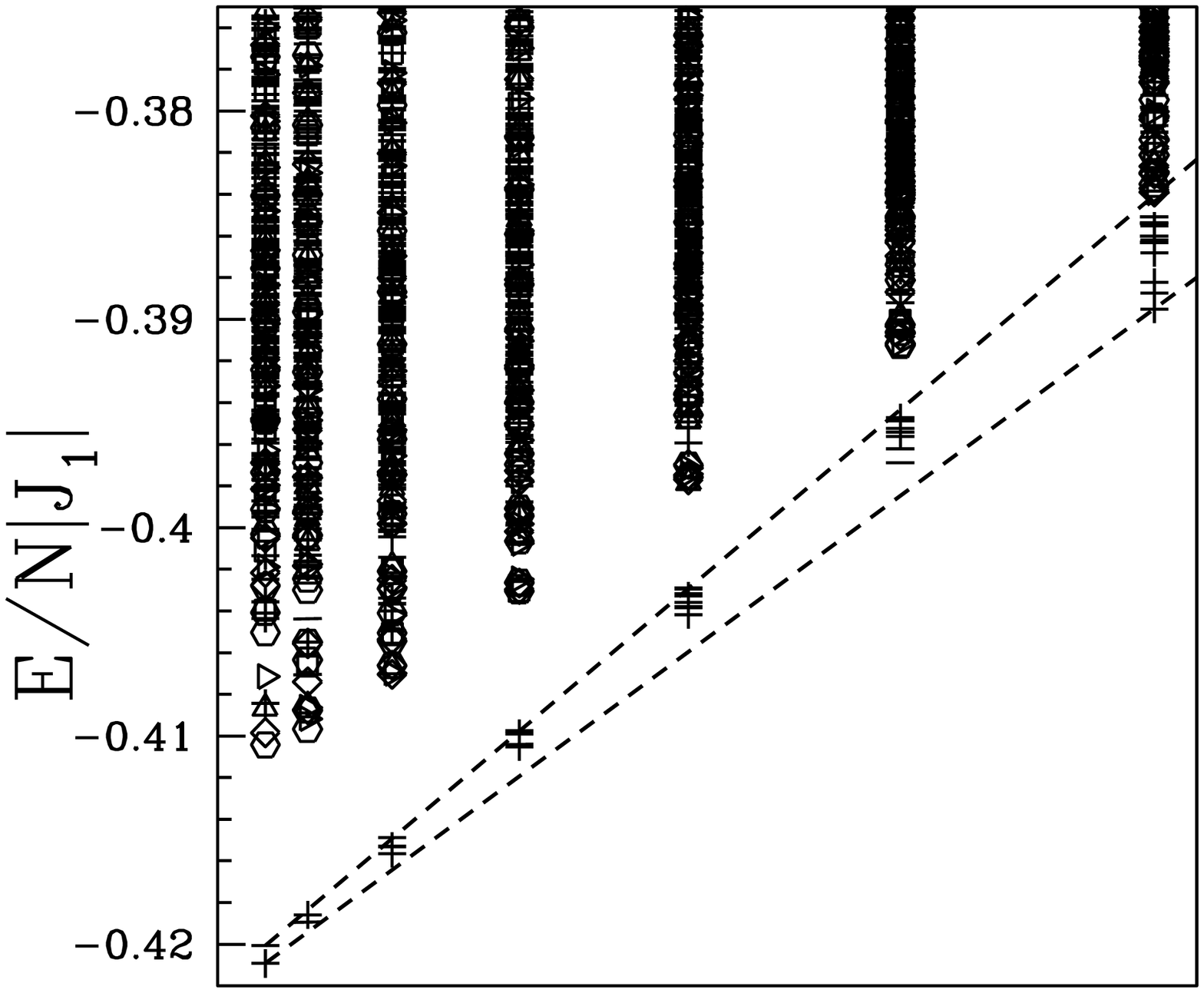}}
\resizebox{.4\textwidth}{!}{\includegraphics*[0.5cm,7cm][20cm,16.5cm]{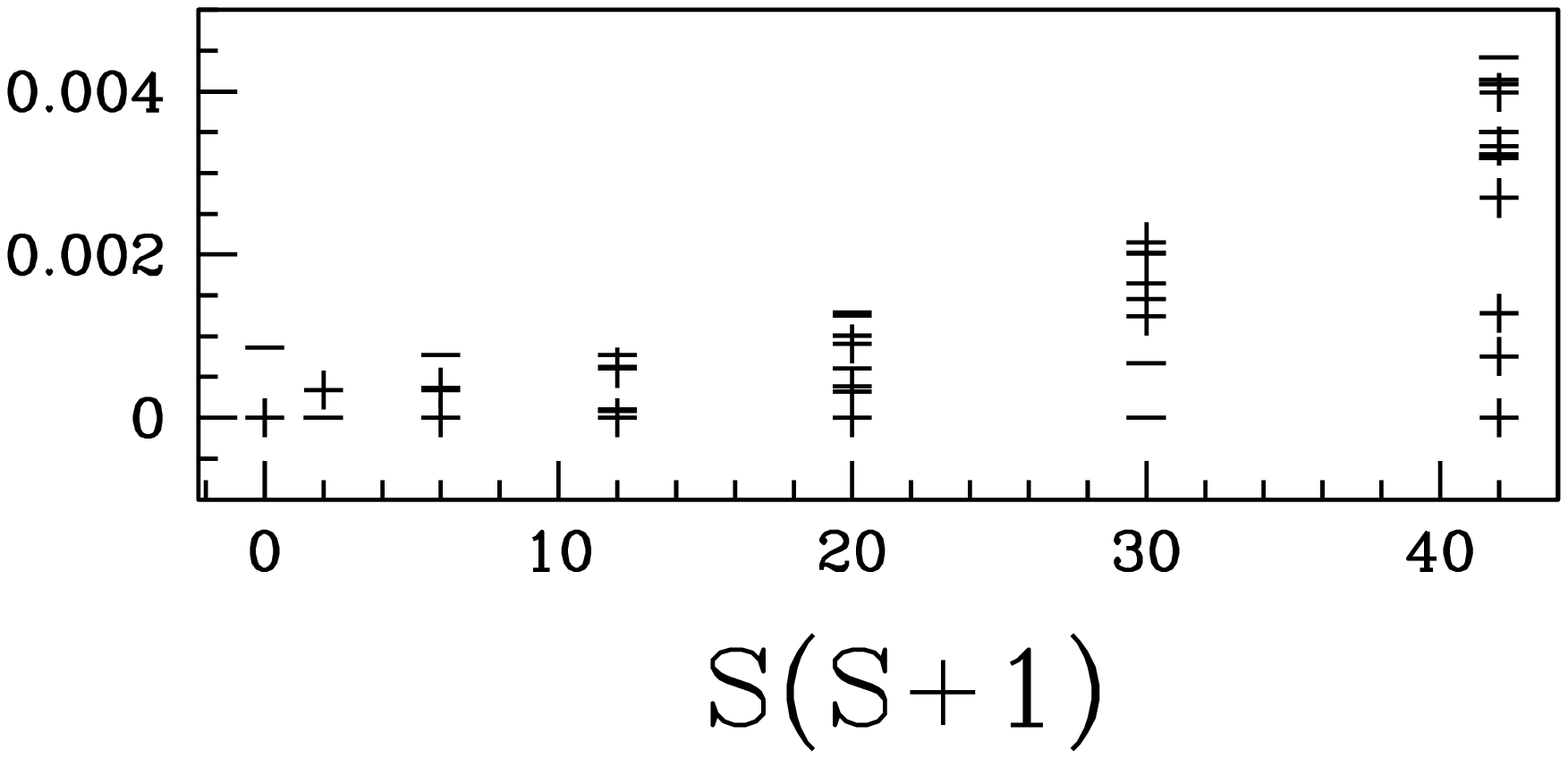}}

\caption{\label{fig:spec} Top:\ Exact spectrum of the 
Hamiltonian~(\ref{H}),\    with  \mbox{$J_2/|J_1|$=0.5},\ for      the
\mbox{$N$=36}   sample shown  in  Fig.~\ref{fig:samples}.\ The exact
energies per  spin   are displayed versus   \mbox{$S(S+1)$}.\  The $+$
(resp. $-$) symbols  are   eigenstates with even (resp.   odd)  parity
under  the $\mathcal{R}_\pi$    lattice rotation.\  Other   symbols  are
eigenstates without the $\mathcal{R}_\pi$ symmetry.\ The full symmetries
of the  lowest eigenstates are  given  in Table~\ref{tab:qdjs36}.\ The
expected QDJS for the 12-sublattice  N\'eel order appear between the two
dashed lines.\ In particular,\ even and odd parity eigenstates come in
an  equal  number in each  spin  sector,\  as expected  for the \Zdeux
symmetry breaking.\ Bottom:\ Zoom on the QDJS (energies rescaled).}
\end{figure}

We  determine the number  and degeneracies of the  QDJS expected for a
sample  with  \mbox{$N$=36}  spins  and   compare this  to  the  exact
diagonalization       result.\\       We  first         compute    the
decomposition~(\ref{dec})  for    total spin     \mbox{$S\leq6$},\    the
approximate   maximum spin of  the QDJS  for \mbox{$N=36$}.\ As stated
earlier,\  we  just have to take  Table~\ref{tab:decex}  and perform the
relevant permutations
\mbox{$A_1\leftrightarrow A_2$}  and \mbox{$T_1\leftrightarrow T_2$}.\   In order to take  the whole
\grOh  group into account we  recall  that each one of   the IRs of the
decomposition  should actually   appear twice with   the  two possible
parities  under the spin flip operation.\ Then  we  map these IRs onto
those of \grGN and here we need to specify  the shape of the sample we
diagonalize (Fig.~\ref{fig:samples}).
The  mapping is  readily done  since  the \mbox{$N$=36} sample has the
full
\csixv symmetry and we may directly use Table~\ref{tab:map}.\
Hence we have obtained the full composition of the Anderson's tower of
states   in     the     lowest   spin    sectors     \mbox{$0\leq S\leq6$}
(Table~\ref{tab:dec36}).\ Again we stress that each QDJS appearing in
Table~\ref{tab:dec36} should be present  twice,\ with the two parities
under the \mbox{$R_\pi$} operation,\ this last result being the hallmark
of the \Zdeux symmetry breaking observed classically.
\begin{center}
\begin{table}[h]
\caption{\label{tab:dec36} Expected symmetries ($\bm{k}$: wave vector,\
$\mathcal{R}_{2\pi/3}$: phase  factor  in a lattice ${2\pi/3}$  rotation,\
and $\sigma$:\ phase factor in a lattice reflection) and dimensionality (d)
of the IRs  appearing in the QDJS  of the 12-sublattice N\'eel state.\
Each of  these  IRs  actually  appears    twice in  the  spectra   with
\mbox{$\mathcal{R}_\pi=\pm1$} (not  shown),\  these two copies  will  be
noted in the following  $e$ (for even)  and $o$  (for odd).\  The last
columns give the number of  each IR expected in  the $S$ sector (up to
$S$=6) for the
\mbox{$N$=36} sample,\ according to Eq.~(\ref{nnus}).\ The 
last line gives the number $n(S)$ of QDJS in each sector [counting the
\Zdeux replica gives a total number of $2\,n(S)$ QDJS in the spin $S$
sector]. \\}
\begin{tabular}{|c||lllrc|c||ccccccc|}\hline
\multicolumn{7}{|c||}{$S$}&0&1&2&3&4&5&6\\\hline\hline
1&$\bm{k}=\bm{0}$&$\mathcal{R}_{2\pi/3}$=1&$\sigma$=&1&&d=1&0&0&0&1&0&0&1\\\hline
2&$\bm{k}=\bm{0}$&$\mathcal{R}_{2\pi/3}$=1&$\sigma$=&-1&&d=1&1&0&0&0&1&0&1\\\hline
3&$\bm{k}=\bm{0}$&$\mathcal{R}_{2\pi/3}$=$j,j^2$&&&&d=2&0&0&1&0&1&1&1\\\hline
4&$\bm{k}=\bm{X}_{1,2,3}$&&$\sigma$=&-1&&d=3&0&0&1&1&1&1&2\\\hline
5&$\bm{k}=\bm{X}_{1,2,3}$&&$\sigma$=&1&&d=3&0&1&0&1&1&2&1\\\hline\hline
\multicolumn{7}{|c||}{$n(S)$}&1&3&5&7&9&11&13\\\hline
\end{tabular}
\end{table}
\end{center}
The  comparison with the  exact spectrum  of the Hamiltonian~(\ref{H})
for the  \mbox{$N$=36} sample  is straightforward (Fig.~\ref{fig:spec}
and TABLE~\ref{tab:qdjs36}).\ Indeed,\  one   clearly notes a  set  of
low-energy eigenstates,\ well  separated  from the magnon  excitations
for    total   spin   \mbox{$S\leq$6},\   which     scale reasonably  as
\mbox{$S(S+1)$}.\ But the  main point is  that for each total spin $S$
the   number and symmetries of  the  low-lying eigenstates are exactly
those obtained from our symmetry analysis,\ as can be readily verified
in TABLE~\ref{tab:qdjs36}.
\begin{center}
\begin{table}[h]
\caption{\label{tab:qdjs36} Energies per spin and symmetries of the 
lowest eigenstates  of the Hamiltonian~(\ref{H}) for $S\leq4$. \\}
\begin{tabular}{|c|l|c|}\hline
$S$&\multicolumn{1}{c|}{$E/N|J_1|$}&IR\\\hline
0&-0.420920&2e\\
0&-0.420054&2o\\\hline
1&-0.418935&5o\\
1&-0.418596&5e\\\hline
2&-0.415654&3e\\
2&-0.415317&4o\\
2&-0.415282&4e\\
2&-0.414885&3o\\\hline
3&-0.410506&1e\\
3&-0.410423&4o\\
3&-0.410403&5o\\
3&-0.409898&4e\\
3&-0.409886&5e\\
3&-0.409738&1o\\\hline
4&-0.404181&3e\\
4&-0.403855&5e\\
4&-0.403797&2e\\
4&-0.403574&4o\\
4&-0.403272&4e\\
4&-0.403172&5o\\
4&-0.402928&2o\\
4&-0.402888&3o\\\hline
\end{tabular}
\end{table}
\end{center}
In particular,\  we note that we  have \mbox{$2(2S+1)$} QDJS in each
total  spin     $S$    sector,\  consistently   with      a   complete
\mbox{\SUdeux breaking} in the thermodynamic limit,\ with the factor 2
coming from the two replica \mbox{$\mathcal{R}_\pi=\pm1$} of each QDJS and
taking care of  the \Zdeux symmetry breaking.\\  The same analysis has
been  made for \mbox{$N$=12}  and 24 nonfrustrating samples,\ leading
to the  same result.\\ These  symmetry arguments are strong  enough to
claim  that,\ at  least in  a   certain range of parameters,\  quantum
fluctuations   do not  destroy    the complicated  \mbox{12-sublattice}
classical long-range order and that there exists a {\it quantum cuboc}
phase in the  thermodynamic  limit,\ in the sense   of a ground  state
consisting of   the classical \mbox{\it  cuboc}  state renormalized by
quantum  fluctuations,\ as explained  earlier.\\  However,\ it  can be
objected that  long-wavelength quantum fluctuations,\  which cannot be
accounted for on the small samples we diagonalized,\ may wipe out the
{\it cuboc} order.
\section{Semiclassical approach}
\label{sec:sw}
\begin{figure}[h]
\begin{center}
\resizebox{.4\textwidth}{!}{\includegraphics*[0cm,0cm][10cm,5.5cm]{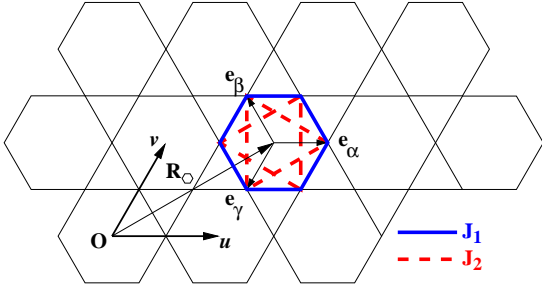}}
\caption{(Color online)\label{fig:latt} Kagom\'e lattice with
nearest (solid) and next-nearest neighbor exchange (dashed).\ Note
that we  span all the exchange paths by considering the 12 links
inside an empty hexagon $\hexagon$ located at $\bm{R}_{\hexagon}$
and then iterate  over the whole Bravais lattice.\ ${\bm u}$ and
${\bm  v}$ are the basis  vectors of the Bravais lattice of length
2 and the three sites \mbox{$\alpha,\beta,\gamma$} per Bravais
cell define the three unit vectors
\mbox{$\bm{e}_\alpha=\bm{u}/2$},\
\mbox{$\bm{e}_\beta=(\bm{v}-\bm{u})/2$} and
\mbox{$\bm{e}_\gamma=-\bm{v}/2$}.}
\end{center}
\end{figure}

\begin{figure}[h]
\begin{center}
\resizebox{.3\textwidth}{!}{\includegraphics*[2.3cm,1.5cm][17.2cm,19cm]{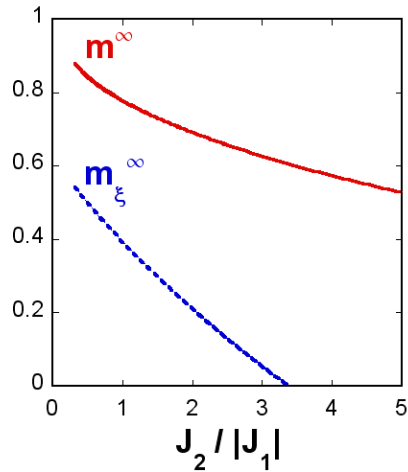}}
\caption{(Color online) \label{fig:sw}Extrapolated values of the magnetization in
the local basis $m^{\infty}$ (solid line) and the alternate scalar
chirality    $m_{\xi}^{\infty}$ (dashed line)  as  a function   of
$J_2/|J_1|$.}
\end{center}
\end{figure}

Now  that we are  convinced  that the  \mbox{12-sublattice} N\'eel order
observed in  the  classical {\it  cuboc}  phase also  exists in  small
samples of spins 1/2,\  we may compute  the effect  of long-wavelength
quantum fluctuations on  the energy,\  sublattice magnetization,\  and
chiral   order parameter  in the  ground    state using the  spin-wave
approximation.\\ The   route to compute   the quantum deviations  to a
classical  {\it cuboc} state is  straightforward:\ at each site of the
kagom\'e lattice we define a local frame in spin space whose $z$ axis is
aligned with the local spin in the  classical ground state.\ Thus,\ in
this frame  the  classical  {\it  cuboc} ground    state  is simply  a
ferromagnetic    state  to  which    one   can  readily   apply    the
Holstein-Primakov transformation.\\ First,\ we choose a particular
\mbox{\it cuboc} state,\ say,\ the one of Fig.~\ref{fig:cuboc}.\ At a
given site $i$ of the kagom\'e lattice we define the local frame
\mbox{($\bm{x}_i$,\ $\bm{y}_i$,\  $\bm{z}_i$)} with  $\bm{z}_i$
the unit  vector parallel to  the local classical spin $\bm{S}_i$.\ To
choose $\bm{x}_i$ we note that each site of the kagom\'e lattice belongs
to two triangles pointing toward opposite  directions.\ Consider the other
two spins on the downward triangle and label them $j$ and $k$,\ with
\mbox{($i$,$j$,$k$)} turning  clockwise.\  The directions of   the two
spins \mbox{$j$,$k$} in our  \mbox{\it cuboc} state are $\bm{z}_j$ and
$\bm{z}_k$ and one can easily verify that
\mbox{$\bm{x}_i=(\bm{z}_k-\bm{z}_j)/\sqrt{2}$} is  indeed a unit vector
orthogonal to $\bm{z}_i$.\ We completely determine the local frame by
imposing \mbox{$\bm{y}_i=\bm{z}_i\land\bm{x}_i$}.\\ This   construction is
translationally invariant and repeating it  for all  the sites of  the
kagom\'e lattice will lead to  12 different local frames associated to
the 12  sublattices of the classical ground  state.\ Hence,\ using the
appropriate transition matrix $\mathcal{R}_i$ from the reference frame
\mbox{($\bm{x}$,$\bm{y}$,$\bm{z}$)} to the local frame
\mbox{($\bm{x}_i$,$\bm{y}_i$,$\bm{z}_i$)},\ we may compute the components 
of the spin at site $i$ in its local frame
\mbox{$\bm{S}'_i=(S_i^{x_i},S_i^{y_i},S_i^{z_i})$} from
\mbox{$\bm{S}_i=\mathcal{R}_i\bm{S}'_i$} (Remark:\ there are only 12 
different   $\mathcal{R}_i$     matrices).\\   Before  computing   the
Hamiltonian~(\ref{H}) in the  local frame,\  we note  that it can   be
rewritten    as    a   sum    over     the    $N/3$   empty   hexagons
(Fig.~\ref{fig:latt}):
\begin{equation}
{\cal H}  = J_1\sum_{\hexagon}\sum_{<i,j>} {\bf  S}_i\cdot{\bf S}_j + J_2\sum_{\hexagon}\sum_{<<i,k>>}
{\bf S}_i\cdot{\bf S}_k,
\label{H2}
\end{equation}
where \mbox{$<i,j>$} and \mbox{$<<i,k>>$} are now,\ respectively,\ the six
nearest and six  next-nearest neighbor pairs of  sites enclosed in the
empty hexagon $\hexagon$.\ Now,\ using
\mbox{$\bm{S}_i\cdot\bm{S}_j=\bm{S}'_iT_{ij}\bm{S}'_j$},\ with
\mbox{$T_{ij}={}^t\mathcal{R}_i\mathcal{R}_j$},\ we may rewrite~(\ref{H2}) in the
local frame as
\begin{equation}
{\cal H} =   J_1\sum_{\hexagon}\sum_{<i,j>}  {\bf  S}'_iT_{ij}{\bf  S}'_j  +
J_2\sum_{\hexagon}\sum_{<<i,k>>} {\bf S}'_iT_{ik}{\bf S}'_k.
\label{H'}
\end{equation}
Now,\ we are able to quantize the fluctuations around the classical
\mbox{\it cuboc} state using \mbox{Holstein-Primakov} bosons.\ Up
to quadratic order the transformation at site $i$ is written as
\begin{equation}
\left\{
\begin{array}{lllllll}
S^{\prime+}_i&=&S_i^{x_i}+jS_i^{y_i}&=&\sqrt{2S-n_i}\,c_i&\simeq&\sqrt{2S}\,c_i,\\
S^{\prime-}_i&=&S_i^{x_i}-jS_i^{y_i}&=&c_i^\dagger\sqrt{2S-n_i}&\simeq&\sqrt{2S}\,c_i^\dagger,\\
S^{\prime z}_i&=&S_i^{z_i} &=&S-n_i & &,
\end{array}\right.
\label{HP}
\end{equation}
where $c_i^\dagger$  and $c_i$,\  respectively,  create and annihilate
a Holstein-Primakov boson at site $i$,\ with $S$ the length of the
classical     local      spin     and     \mbox{$n_i=c_i^\dagger    c_i$}.\\
Inserting~(\ref{HP}) into~(\ref{H'}) we obtain  the quantized version of
the original Hamiltonian~(\ref{H}) up to quadratic order.\ As usual we
Fourier-transform $c_i$ and $c_i^\dagger$ according to
\begin{equation}
c_i=\sqrt{\frac{3}{N}}\sum_{\bm q}e^{-j\bm{q}\cdot(\bm{R}_{\hexagon}+\bm{e}_{\mu_i})}c_{\bm q}^{\mu_i},
\end{equation}
where the sum runs over the first Brillouin zone,\ $\bm{R}_{\hexagon}$
is  a vector of the  Bravais  lattice,\ and \mbox{$\mu_i=\alpha,\beta,\gamma$} indicates
one    of the     three  possible    sites   in   the   Bravais   cell
(Fig.~\ref{fig:latt}).\\  Again,\ in the  local  frame the complicated
12 sublattice order is  just  a ferromagnetic  state,\ so that  we
need only three flavors of bosons
\mbox{$\mu_i=\alpha,\beta,\gamma$} associated with the  three sites per Bravais cell  on
the kagom\'e lattice.
\ We may finally bring the Hamiltonian to matrix form as
\begin{equation}
{\cal H}=(J_1-J_2)NS(S+1)+\sum_{\bm q}V_{\bm q}^\dagger M_{\bm q}V_{\bm q},
\label{Hq}
\end{equation}
where the sum runs over the entire first  Brillouin zone,\ $V_{\bm q}$
is the column  vector  \mbox{($c_{\bm q}^\alpha$,\ $c_{\bm q}^\beta$,\  $c_{\bm
q}^\gamma$,\   $c_{-{\bm   q}}^{\alpha\dagger}$,\ $c_{-{\bm   q}}^{\beta\dagger}$,\    $c_{-{\bm
q}}^{\gamma\dagger}$)},\ and $M_{\bm q}$ is the
\mbox{$6\times6$}  matrix \mbox{$M_{\bm    q}=(J_2-J_1)S\mathbbm{1}+\left(\begin{array}{cc}A_{\bm
q}&B_{\bm q}\\B_{-{\bm    q}}&A_{-{\bm    q}}\end{array}\right)$},\ with
$\mathbbm{1}$  the identity  matrix,\ and
\begin{equation}
A_{\bm q}=\left(\begin{array}{ccc}0&a_{\alpha\beta}({\bm q})&a_{\alpha\gamma}({\bm q})\\a_{\alpha\beta}({\bm q})&0&a_{\beta\gamma}({\bm q})\\a_{\alpha\gamma}({\bm q})&a_{\beta\gamma}({\bm q})&0\end{array}\right),
\end{equation}
and
\begin{equation}
B_{\bm q}=\left(\begin{array}{ccc}0&b_{\alpha\beta}({\bm q})&b_{\alpha\gamma}({\bm q})\\b_{\alpha\beta}(-{\bm q})&0&b_{\beta\gamma}({\bm q})\\b_{\alpha\gamma}(-{\bm q})&b_{\beta\gamma}(-{\bm q})&0\end{array}\right),
\end{equation}
with
\begin{eqnarray*}
a_{\alpha\beta}({\bm     q})&=&\frac{J_2S}{4}\cos    q_{\beta-\alpha}-\frac{J_1S}{4}\cos
q_\gamma-\frac{J_1S}{\sqrt{2}}\sin           q_\gamma,\\
b_{\alpha\beta}({\bm     q})&=&\frac{J_2S}{4}\cos    q_{\beta-\alpha}-\frac{J_1S}{4}\cos
q_\gamma+\frac{J_2S}{\sqrt{2}}\sin           q_{\beta-\alpha},
\end{eqnarray*}
with the four remaining matrix elements obtained by cyclic permutation
of  the  indices   and  where we  have    used the condensed  notation
$q_{\beta-\alpha}={\bm q}\cdot({\bm e}_\beta-{\bm e}_\alpha)$.\\  Note  that the first  term
in~(\ref{Hq})   contains the usual     dominant  contribution to   the
renormalization of the classical ground state energy
\mbox{$(J_1-J_2)NS^2$}.\ Note also  that $A_{\bm  q}^\dagger=A_{\bm q}$  and
$B_{\bm q}^\dagger=B_{-{\bm q}}$ so  that $M_{\bm q}$ is  indeed hermitic.\\
As usual,\ the next step is to find a transition matrix $P_{\bm q}$ such
as $M_{\bm q}$ is diagonal in the new basis.\\ It should be emphasized
that  $P_{\bm q}$ is  strongly prescribed by the fact  that  it has to
preserve the boson commutation relations~\cite{c78},\ much in the same
way as the Bogoliubov tranformation for the collinear antiferromagnet
on  the square lattice.\\  We end up  with  three Bogoliubov bosons,\
obtained from
\mbox{$W_{\bm  q}$=($d_{\bm q}^\alpha$,\ $d_{\bm  q}^\beta$,\  $d_{\bm  q}^\gamma$,\
$d_{-{\bm         q}}^{\alpha\dagger}$,\   $d_{-{\bm  q}}^{\beta\dagger}$,\       $d_{-{\bm
q}}^{\gamma\dagger}$)=$P_{\bm q}V_{\bm q}$},\   and the six eigenvalues  give the
corresponding three dispersion branches.\\ In the new basis (\ref{Hq})
reads as
\begin{equation}
{\cal     H}=(J_1-J_2)NS(S+1)+\sum_{\bm   q}\sum_{\mu=\alpha,\beta,
\gamma}\omega_{\bm   q}^\mu(d_{\bm q}^{\mu\dagger}d_{\bm q}^\mu+\frac{1}{2}),
\label{Hsw}
\end{equation}
so that the energy per spin in the ground state,\ which is the
vacuum \ket{0} of $d_{\bm q}^\mu$ Bogoliubov bosons,\  is simply
\begin{equation}
e_0^N=\frac{1}{N}\langle0|{\cal
H}|0\rangle=(J_1-J_2)S(S+1)+\frac{1}{N}\sum_{\bm
q}\sum_{\mu=\alpha,\beta,\gamma}\frac{\omega_{\bm q}^\mu}{2}.\label{esw}
\end{equation}
As  for  the  collinear antiferromagnet,\  note   that the Bogoliubov
transformation $P_{\bm q}$ is actually singular wherever \mbox{$\omega_{\bm
q}^\mu=0$},\  namely at  \mbox{$\bm{q}=\bm{X}_{1,2,3}$},\ where soft modes
are expected in the thermodynamic limit.\ One indeed verifies that the
lowest branch vanishes at each  one of the  three edge centers of  the
first  Brillouin  zone  giving exactly three   Goldstone  modes in the
thermodynamic  limit,\ as  expected  for a  complete  \SUdeux symmetry
breaking.\\   We   may   then compute  the    renormalization   of the
magnetization in the local basis,\ 
\begin{equation}
m^N=\frac{1}{NS}\langle0|\sum_{i=1}^NS_i^{z_i}|0\rangle=1+\frac{1}{S}(1-\frac{1}{N}\sum_{\bm
q}{}^{'}\sum_{i,j=1}^3(P_{\bm q}^{i,j})^2), \label{msw}
\end{equation}
where the prime denotes  a sum over  the first Brillouin zone deprived
of its three edge centers and where \mbox{$P_{\bm q}^{i,j}$} is the
\mbox{$(i,j)$th} matrix element of the matrix $P_{\bm q}$.\\ Another quantity of
interest to us  is the  renormalization  of the scalar chirality  on a
triangle.\  It is naturally normalized by   its value in the classical
ground state,\ so that we define
\mbox{$\xi_\bigtriangleup=\frac{\sqrt{2}}{S^3}(\bm{S}_i\land\bm{S}_j)\cdot\bm{S}_k$} on each triangle
\mbox{($i$,$j$,$k$)},\ with \mbox{($i$,$j$,$k$)} turning clockwise.\
We    may then compute the   renormalization   of the alternate scalar
chirality in the ground state as
\begin{equation}
m_\xi^N=\frac{3}{2N}\langle0|\sum_\bigtriangleup(-1)^{\alpha_\bigtriangleup}\xi_\bigtriangleup|0\rangle=1+\frac{3}{S}(1-\frac{1}{N}\sum_{\bm
q}{}^{'}\mathcal{D}_{\bm q}), \label{mssw}
\end{equation}
where the  sum runs over the $2N/3$  triangles of the kagom\'e lattice,\
while $\alpha_\bigtriangleup$ is,\  respectively,\ 0 and 1  on upward and downward triangles,\ 
and
\begin{equation}
\begin{array}{rl}
\mathcal{D}_{\bm q}=
\displaystyle{\sum_{i,j=1}^3}&(P_{\bm q}^{i,j})^2\\
-\frac{1}{2}\displaystyle{\sum_{j=1}^3}&\left(
\begin{array}{ll}
&\cos  q_\alpha(P_{\bm    q}^{2,j}+P_{\bm  q}^{5,j})(P_{\bm q}^{3,j}+P_{\bm
q}^{6,j})\\\\ +&\cos q_\beta(P_{\bm     q}^{1,j}+P_{\bm   q}^{4,j})(P_{\bm
q}^{3,j}+P_{\bm   q}^{6,j})\\\\ +&\cos  q_\gamma(P_{\bm     q}^{1,j}+P_{\bm
q}^{4,j})(P_{\bm             q}^{2,j}+P_{\bm             q}^{5,j})\\\\
+&\frac{\sqrt{2}}{3}\sin   q_\alpha(P_{\bm   q}^{5,j}P_{\bm q}^{6,j}-P_{\bm
q}^{2,j}P_{\bm    q}^{3,j})\\\\  +&\frac{\sqrt{2}}{3}\sin   q_\beta(P_{\bm
q}^{4,j}P_{\bm     q}^{6,j}-P_{\bm      q}^{1,j}P_{\bm   q}^{3,j})\\\\
+&\frac{\sqrt{2}}{3}\sin   q_\gamma(P_{\bm q}^{4,j}P_{\bm   q}^{5,j}-P_{\bm
q}^{1,j}P_{\bm q}^{2,j})
\end{array}\right).
\end{array}
\end{equation}
These quantities were numerically computed on finite size samples
with linear sizes $L\leq10^2$ Bravais lattice spacings.\  As usual,\ 
the leading correction   to  the  classical value  comes   from
the  first magnon excitation whose energy scales as $1/L$.\ We
find very good agreement with  the   expected scaling  laws  for
$e_0^N$ \mbox{($\sim1/L^3$)},\ $m^N$
\mbox{($\sim1/L$)},\ and $m_\xi^N$ \mbox{($\sim1/L$)},\ and
perform the extrapolation to  the thermodynamic limit.\\ We first note
in  (Fig.~\ref{fig:sw}) that in  the  thermodynamic limit $m^{\infty}$  and
$m_\xi^{\infty}$ remain finite in a large  region around the point were exact
diagonalizations  were performed,\  thus showing the  stability of the
{\it cuboc}  phase against long-wavelength quantum fluctuations.\ For
\mbox{$J_2/|J_1|=0.5$} we find   to  lowest  order in  the  spin  wave
approximation that $m^{\infty}$ and $m_\xi^{\infty}$ are renormalized by $16\%$ and
$50\%$,\   respectively.\\   However,\  upon   increasing    $J_2$,\ both
quantities  decrease drastically.  For  \mbox{$J_2/|J_1|\gsim  3$} the
chirality disappears  which casts a strong doubt  on  the stability of
the 12-sublattice state itself in this range of parameters.
\section{A gapped phase for \mbox{$\bm{J_2/|J_1|=5.0}$}}
\label{sec:gappedphase} 
\begin{figure}[h]
\begin{center}
\resizebox{.4\textwidth}{!}{\includegraphics*[0.5cm,5.5cm][20cm,24.2cm]{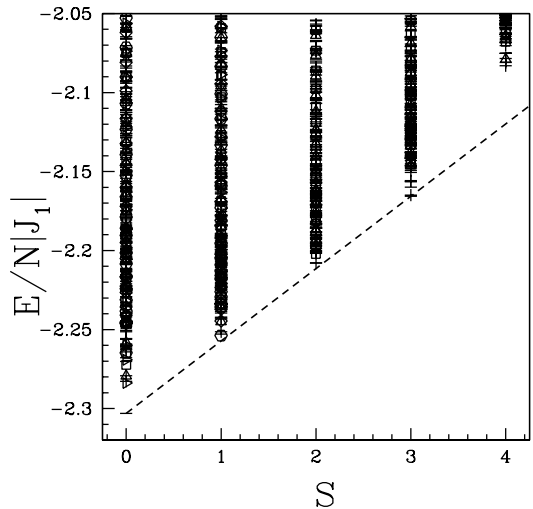}}
\caption{\label{fig:specgap} An exact spectrum of the \mbox{$N$=36} sample for 
\mbox{$J_2/|J_1|$=5.0} versus total spin $S$.\ The symbol convention  is the same as
that of Fig.~\ref{fig:spec}.}
\end{center}
\end{figure}
\begin{figure}[h]
\begin{center}
\parbox{.23\textwidth}{\resizebox{.23\textwidth}{!}{\includegraphics*[0.2cm,0cm][17.5cm,17cm]{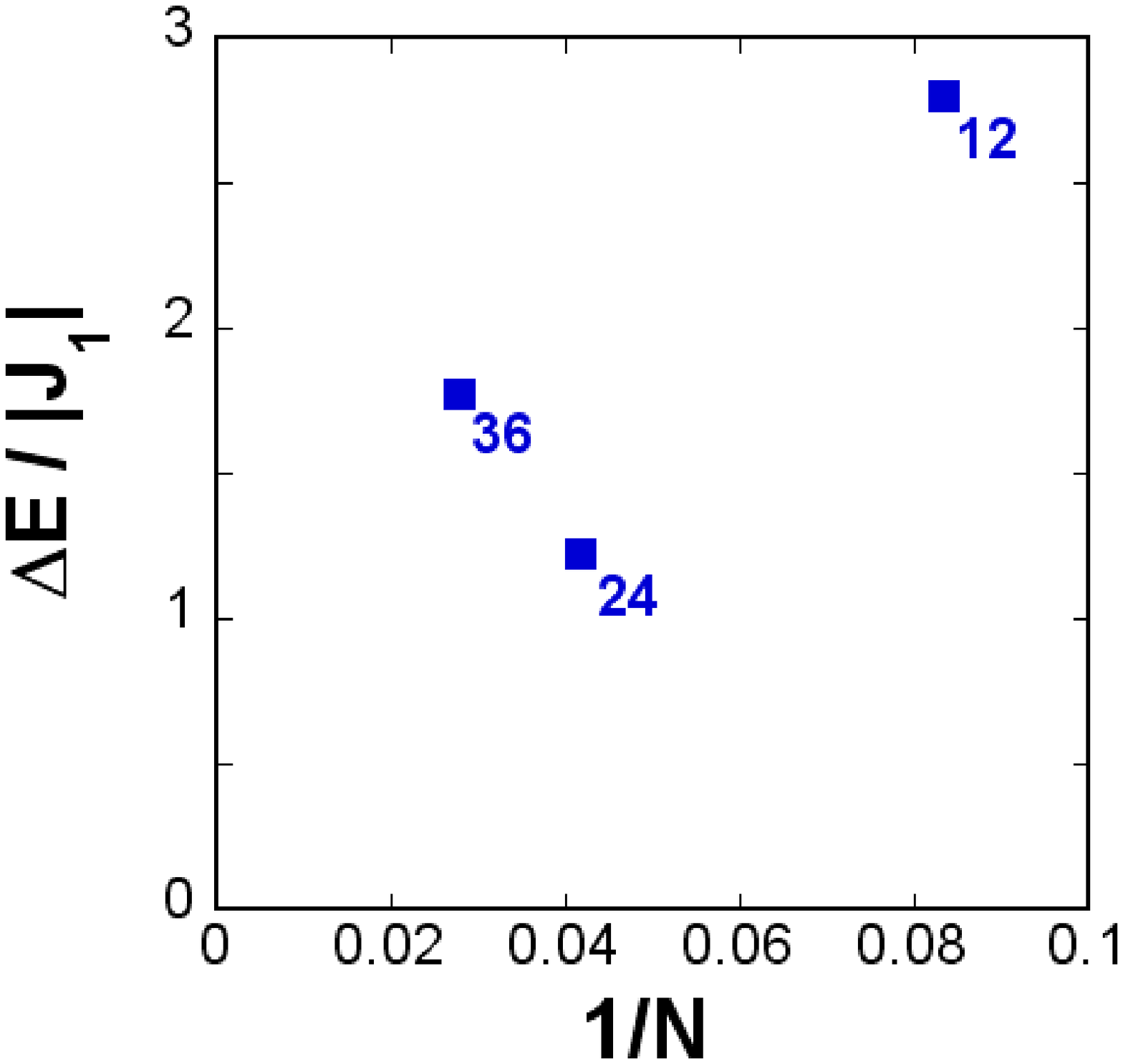}}}
\parbox{.23\textwidth}{\resizebox{.23\textwidth}{!}{\includegraphics*[0.2cm,0cm][17.5cm,17cm]{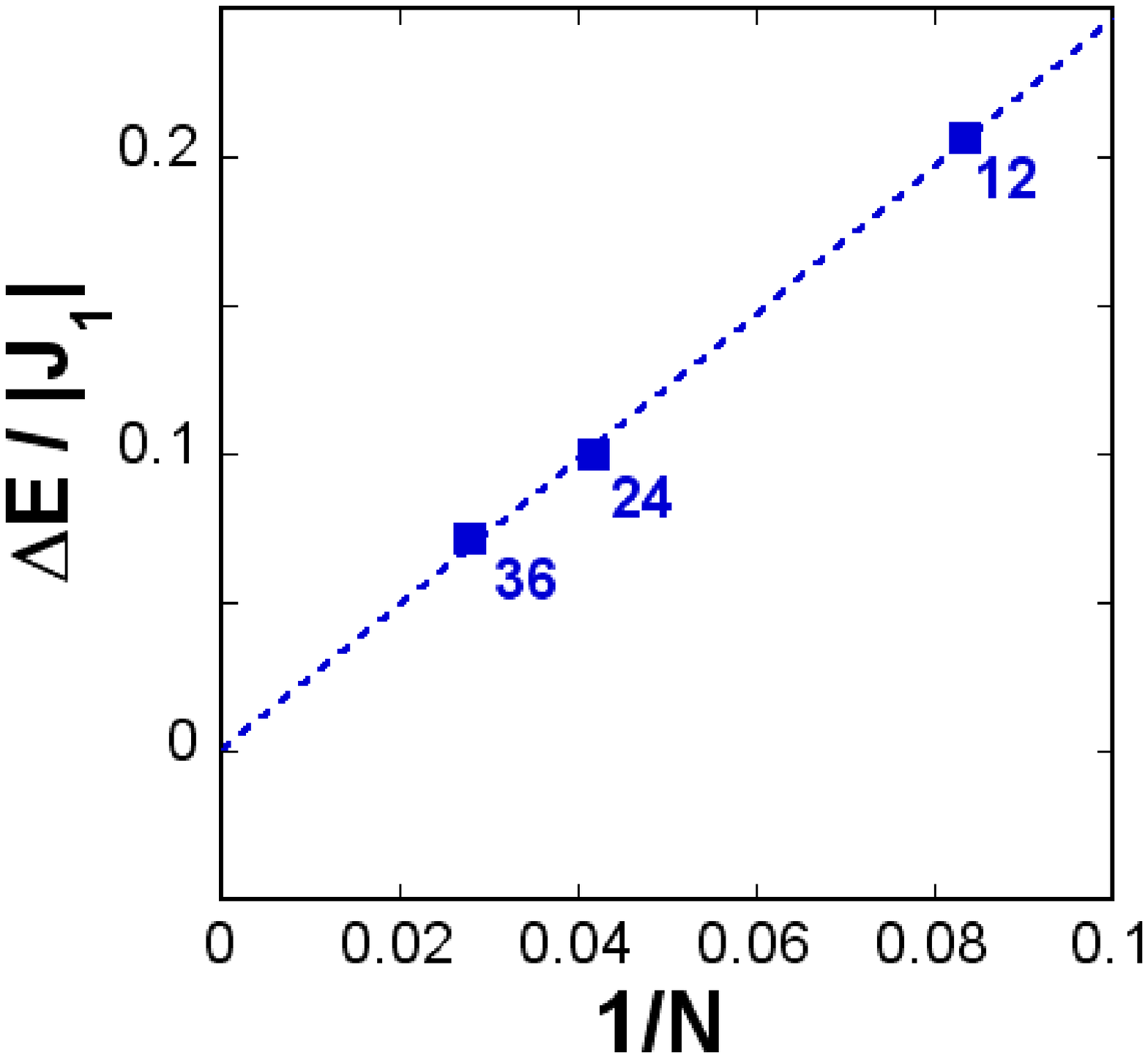}}}
\caption{(Color online) \label{fig:gap} Left:\ Finite size scaling of the spin gap $\Delta E$ 
for $J_2/|J_1|=5.0$, in the supposed to be gapped phase.\ Right:\ The
same gap in the 12-sublattice N\'eel  phase,\ for $J_2/|J_1|=0.5$,\ closes
as $1/N$,\  as expected for the  \`a la  N\'eel \SUdeux symmetry breaking
(Sec.~\ref{subsec:alaneel}).}
\end{center}
\end{figure}

Exact spectra  for \mbox{$J_2/|J_1|=5.0$}  differ notably from spectra
of an ordered phase as  described in Sec.~\ref{sec:quantum}.\ First of
all,\ their  low-lying levels   in each $S$   sector do  not scale  as
$S(S+1)$    but    rather    as  $S$,\     as    can   be    seen   in
Fig.~\ref{fig:specgap}.\   Second,\ the  gap  to  the first excitation
seemingly  does not  close to  zero with the   system size,\ as shown in
Fig.~\ref{fig:gap}.\ This second result  is consistent with  the first
one: in a  gapped  phase a finite  magnetic  field $H_c$ is needed  to
close the gap.\  This critical field is  directly proportional to  the
first  derivative of the energy versus  $S$.\ For this  given model in
this range  of parameters,\ the available sizes  of exact  spectra are
large enough to infirm the presence of a 12 sublattice N\'eel order
and confirm  the  existence of a gapped  phase.\  However they are too
small to discriminate between  a true  spin  liquid or a  valence bond
crystal,\ and to decide if there is one or two different gapped phases
in this range of parameter.
\section{Conclusion}
\label{sec:ccl} In  this paper we have  studied the
\mbox{$J_1-J_2$} model on the kagom\'e lattice with
\mbox{$J_1<0$} and \mbox{$J_2 >0$}.\  We have found a
\mbox{12-sublattice} ordered phase for \mbox{$J_1<0$} and
\mbox{$J_2/|J_1|>1/3$}.\ This new phase was shown to resist
quantum fluctuations.\ On the exact spectra of small size samples we
found the complete signature of  this complicated N\'eel order,\ \ie the
number and symmetries of the QDJS in the  tower of states,\ based on a
very general group-theoretical approach.\ Moreover,\  in the spin wave
approximation,\ long-wavelength  quantum  fluctuations were   found to
renormalize the order parameter to a finite value in a finite range of
parameters up to  \mbox{$J_2/|J_1| \simeq 3$}.\\ The noncoplanarity of the
12 sublattices  in the {\it cuboc} phase  was shown to induce a chiral
symmetry breaking,\ to which we  associated a chiral order parameter.\
Classically,\  we showed  that this \Zdeux   symmetry was  restored at
finite  temperature,\  consistently  with the Mermin-Wagner  theorem,\
though the exact nature of the transition remains to be investigated.\
We were   also able to  find  the signature of  this discrete symmetry
breaking  on exact spectra.\\  Finally,\  we  showed in the  spin-wave
approach that  the   12  sublattice order   is wiped   out  by quantum
fluctuations  for  \mbox{$J_2/|J_1| \gsim 3$}.\ Exact diagonalizations
indeed confirm the existence of  a  spin-gap phase,\ with short-range
order in spin-spin correlations.\ For this model the largest available
sizes ($N=36$) are too small to give more information on the nature of
this quantum phase. \\ The Laboratoire de Physique Th\'eorique de la Mati\`ere
Condens\'ee is UMR 7600 of the CNRS.
\bibliography{jc}
\end{document}